\documentclass[conference]{IEEEtran}
\IEEEoverridecommandlockouts
\usepackage{cite}
\usepackage[utf8]{inputenc}
\usepackage{amsmath,amssymb,amsfonts}
\usepackage{algorithmic}
\usepackage{graphicx}
\usepackage{textcomp}
\usepackage{xcolor}
\usepackage{outlines}
\usepackage{xspace}
\usepackage{listings}
\usepackage{enumitem}
\usepackage{hyperref}
\usepackage{amsthm}
\usepackage[normalem]{ulem}
\usepackage{booktabs}
\usepackage{pifont}
\usepackage{tikz}
\usepackage{pgfplots}
\pgfplotsset{width=7.5cm,compat=1.12}
\usepgfplotslibrary{fillbetween}

\usepackage{todonotes} 
\usepackage[group-separator={,}]{siunitx}
\newcounter{todocounter}
\newcommand{\numberedtodo}[2]{\stepcounter{todocounter}\todo[inline,color=orange!20!white]{\textbf{TODO~\thetodocounter, #1:} #2}}

\newcommand{\fixup}[1]{{\color{red}\textbf{#1}}}
\DeclareMathOperator*{\argmax}{argmax}
\newcommand{\comment}[1]{}
\newcommand{\configuration}[1]{$\mathbb{C}_{#1}$}
\newcommand{\baseline}[0]{$\mathbb{C}_{1234}$}

\newcommand{\cmark}{\ding{51}}%
\newcommand{\xmark}{\ding{56}}%

\newcommand{\corpusSize}[0]{\num{417} million}
\newcommand{\corpusProjs}[0]{\num{6541}}
\newcommand{\statsDBSize}[0]{four thousand}
\newcommand{\baselineTP}[0]{\num{154}}
\newcommand{\baselinePrec}[0]{\num{67}\%}
\newcommand{\ccOnlyTP}[0]{79}
\newcommand{\scOnlyTP}[0]{89}
\newcommand{\overlapTP}[0]{15}
\newcommand{\warningsTriaged}[0]{859}
\newcommand{\triagedTPs}[0]{183}
\newcommand{\triagedFPs}[0]{676}

\newlist{questions}{enumerate}{1}
\setlist[questions,1]{label=R\arabic*.,ref=R\arabic*}

\graphicspath{ {./images/} }

\theoremstyle{definition}

\lstdefinestyle{Cstyle}{
    numberstyle=\tiny,
    basicstyle=\ttfamily\footnotesize,
    breakatwhitespace=false,
    breaklines=true,
    captionpos=b,
    keepspaces=true,
    numbers=left,
    numbersep=7pt,
    showspaces=false,
    showstringspaces=false,
    showtabs=false,
    tabsize=2,
    xleftmargin=2em,
    framexleftmargin=1.5em,
}

\newcommand\projectname[0]{{\textsc{SwapD}}\xspace}

\begin{document}

\title{Out of Sight, Out of Place: Detecting and Assessing Swapped Arguments}
\author{
    \IEEEauthorblockN{Roger Scott\IEEEauthorrefmark{1}\thanks{\IEEEauthorrefmark{1}Roger Scott performed this work while at GrammaTech.}, Joseph Ranieri\IEEEauthorrefmark{2}, Lucja Kot\IEEEauthorrefmark{2}, Vineeth Kashyap\IEEEauthorrefmark{2}}
    \IEEEauthorblockA{\IEEEauthorrefmark{2}GrammaTech, Inc.
    \\
    \IEEEauthorrefmark{1}rhscott@comcast.net, \IEEEauthorrefmark{2}\{jranieri, lkot, vkashyap\}@grammatech.com}
}

\maketitle

\begin{abstract}
Programmers often add meaningful information about program semantics when naming program entities such as variables, functions, and macros. 
However, static analysis tools typically discount this information when they look for bugs in a program.
In this work, we describe the design and implementation of a static analysis checker called \projectname{}, which uses the natural language information in programs to warn about mistakenly-swapped arguments at call sites. 
\projectname{} combines two independent detection strategies to improve the effectiveness of the overall checker.
We present the results of a comprehensive evaluation of \projectname{} over a large corpus of C and C++ programs totaling \corpusSize{} lines of code.
In this evaluation, \projectname{} found \baselineTP{} manually-vetted real-world cases of mistakenly-swapped arguments, suggesting that such errors---while not pervasive in released code---are a real problem and a worthwhile target for static analysis.
\end{abstract}


\section{Introduction}
\label{sec:intro}

Static analysis tools consist of automated ``checkers'', each of which identifies potential problems by looking for matches of a known code defect pattern or violations of an established program development rule. However, traditional static analysis techniques---such as those based on data-flow analysis---do not use the rich natural language information in programs: variable names, field names in a structure or a class, function names, macro names, etc.
Programmers seldom choose these names at random; they select names that convey information about the semantic concepts they are manipulating, with identifiable patterns in the creation, composition, and usage of those names.
As we show in this work, static analysis tools can and should use these patterns to detect more bugs. 

In this paper we introduce \projectname{}, an automated static analysis checker that uses natural language information to detect mistakenly swapped arguments at call sites.
\autoref{bug:sigkill} shows an example\footnote{All the bugs listed in this paper were found with \projectname{} on real-world code not written by the authors. For the sake of presentation, the listings simplify or elide the code context.} of such a mistaken swap, found with \projectname{} in the open-source code for the editor xvile~\cite{xvile}.
Here, the \texttt{kill} function from \texttt{signal.h} is called incorrectly: the arguments for process identifier and signal have been swapped. 
Because the two arguments are type compatible, even the compiler is unlikely to complain about the swap. 

Incorrect argument ordering is an easy mistake to make when programming in a language that supports positional arguments\footnote{That is, the position of the arguments in a function call denotes which parameter they correspond to.}, especially if the declaration for the callee function is not readily available.
Programmer confusion may be exacerbated by certain function and interface design choices, such as counter-intuitive argument ordering or long parameter lists. In typed programming languages, type checking may catch some swapped argument errors, but not all, as seen in~\autoref{bug:sigkill}. 

\begin{lstlisting}[language=C,escapechar=\#,caption={Bug found with \projectname{}: the arguments on line \texttt{5} are\\ mistakenly swapped.},label=bug:sigkill]
// declaration in signal.h  
int kill(pid_t #\dashuline{pid}#, int #\uwave{sig}#); 

// use in xvile
if (child < 0 && errno == EINTR) {
  kill(#\uwave{SIGKILL}#, #\dashuline{cpid}#);
\end{lstlisting}

Underpinning \projectname{} are two observations about developer behavior when naming program entities. First, programmers often choose argument names that are similar to parameter names, due to an underlying conceptual match between the two~\cite{Liu2016}.
Therefore, as in Listing \ref{bug:sigkill}, an accidental swap may have taken place if both (a) argument names do not \emph{cover} (i.e., have a sufficient correspondence with) their  parameter names, and (b) they would cover if argument positions were swapped. 
Second, if we examine several calls to a function (e.g., calls to a library function in a large code corpus), we find discernible statistical patterns with respect to argument names and their positions in the calls.
Statistically, if argument names are atypical in their current positions, but common in swapped positions, it may indicate an error.

\begin{lstlisting}[language=C,escapechar=\#, caption=Bug found by \projectname{} when parameter names are not\\ available in the declaration.,label=bug:statistical]
// declaration in X11/Xlib.h
extern Bool XQueryExtension(Display, _Xconst char*, int*, int* /* #\dashuline{first\_event\_return}# */, int* /* #\uwave{first\_error\_return}# */);

// use in gpaste
if (XQueryExtension (display, "XInputExtension", &xinput_opcode, #\uwave{\&xinput\_error\_base}#, #\dashuline{\&xinput\_event\_base}#)) { /* ... */
\end{lstlisting}

We have found that these two detection strategies are most effective when used in combination. In particular, we make use of statistical data to reduce both false positives (\S \ref{subsec:vetting}) and false negatives (\S \ref{subsec:statistical}).
As a motivating example, consider the GPaste~\cite{GPaste} bug in \autoref{bug:statistical}, also found with \projectname{}.
Here, no parameter names are available in the declaration, so the second, statistical technique was key to detecting the bug. 

A key feature of \projectname{} is that we split parameter and argument names into smaller units, called morphemes\footnote{From the linguistic term for a unit of meaning in a natural language.}, before applying our techniques.
Operating on morphemes rather than whole names is one of the factors that distinguishes \projectname{} from closely related works~\cite{Rice2017,Pradel2018}. 
Splitting is motivated by the intuition that program identifiers are often constructed by agglutinating two or more morphemes.
\autoref{bug:sigkill} and \autoref{bug:statistical} represent individual examples of this naming behavior. 
Indeed, \autoref{fig:morphemes-corpus} (\S \ref{subsec:morpheme-reasoning}) shows that a significant portion of the corpus uses names containing multiple morphemes.
For example, in \autoref{bug:sigkill}, a na\"ive attempt to match parameters and arguments based on string edit distances could fail, but splitting \texttt{SIGKILL} into \texttt{sig} and \texttt{kill}, and \texttt{cpid} into \texttt{c} and \texttt{pid} makes the correspondence clear.
Our morpheme-based approach also allows us to improve signal by removing morphemes that appear in multiple arguments in a call; those likely represent conceptual information about the calling context rather than about the intended correspondence to parameters.
For example, in \autoref{bug:statistical}, splitting \texttt{xinput\_error\_base} and \texttt{xinput\_event\_base} into constituent morphemes (and eliminating the common morphemes \texttt{xinput} and \texttt{base}) helps identify the underlying pattern---that \texttt{error} and \texttt{event} are statistically more likely in their swapped positions.
We have found real bugs with multi-morpheme names (e.g., \autoref{bug:removeHighCoverageNodes}), as summarized in~\autoref{fig:morpheme-size} (\S \ref{subsec:morpheme-reasoning}). 

The techniques used in \projectname{} are largely programming language agnostic. They are broadly applicable to programs in languages that support positional arguments. We have implemented a \projectname{} prototype targeting C/C++ code; those languages are heavily used in security and reliability-critical software, which provides particular motivation for accurate bug detection. During our empirical evaluation and triage of \projectname{} warnings, we found that many apparent argument swaps are intentional. Thus, we have designed and adapted a variety of techniques to reduce the number of false positives.

Our major contributions include:
\begin{itemize}
  \item A cover-based checker for detecting swapped arguments via mismatches between argument and parameter names.
  \item A statistical checker for swapped arguments based on data collected from a large code corpus.
  \item A morpheme-oriented handling of names in both these approaches, for increasing the relevant signal present in names.
  \item A hybrid approach combining the two checkers and further false positive reduction techniques to achieve high accuracy in detecting swapped-argument errors.
  \item A comprehensive evaluation of \projectname{} on \corpusSize{} lines of open-source C/C++ code corpus~\cite{FedoraSRPM}; we believe our evaluation to be one of the largest in this research area, especially for C/C++. \projectname{} found \baselineTP{} swapped argument errors across this corpus. This figure suggests that, while swapped-argument errors are not extremely common, they are a real problem, and efforts to detect them are likely to provide value to developers. 
\end{itemize} 

The remainder of this paper is organized as follows: we give an overview of \projectname{} (\S\ref{sec:overview}); provide details of specific algorithms and techniques (\S\ref{label:design}); present the results of our empirical evaluation (\S\ref{sec:eval}); discuss related work (\S\ref{sec:related}); and conclude (\S\ref{sec:conclusion}).

\section{Overview}
\label{sec:overview}

\begin{figure*}
\centering

\includegraphics[scale=0.65]{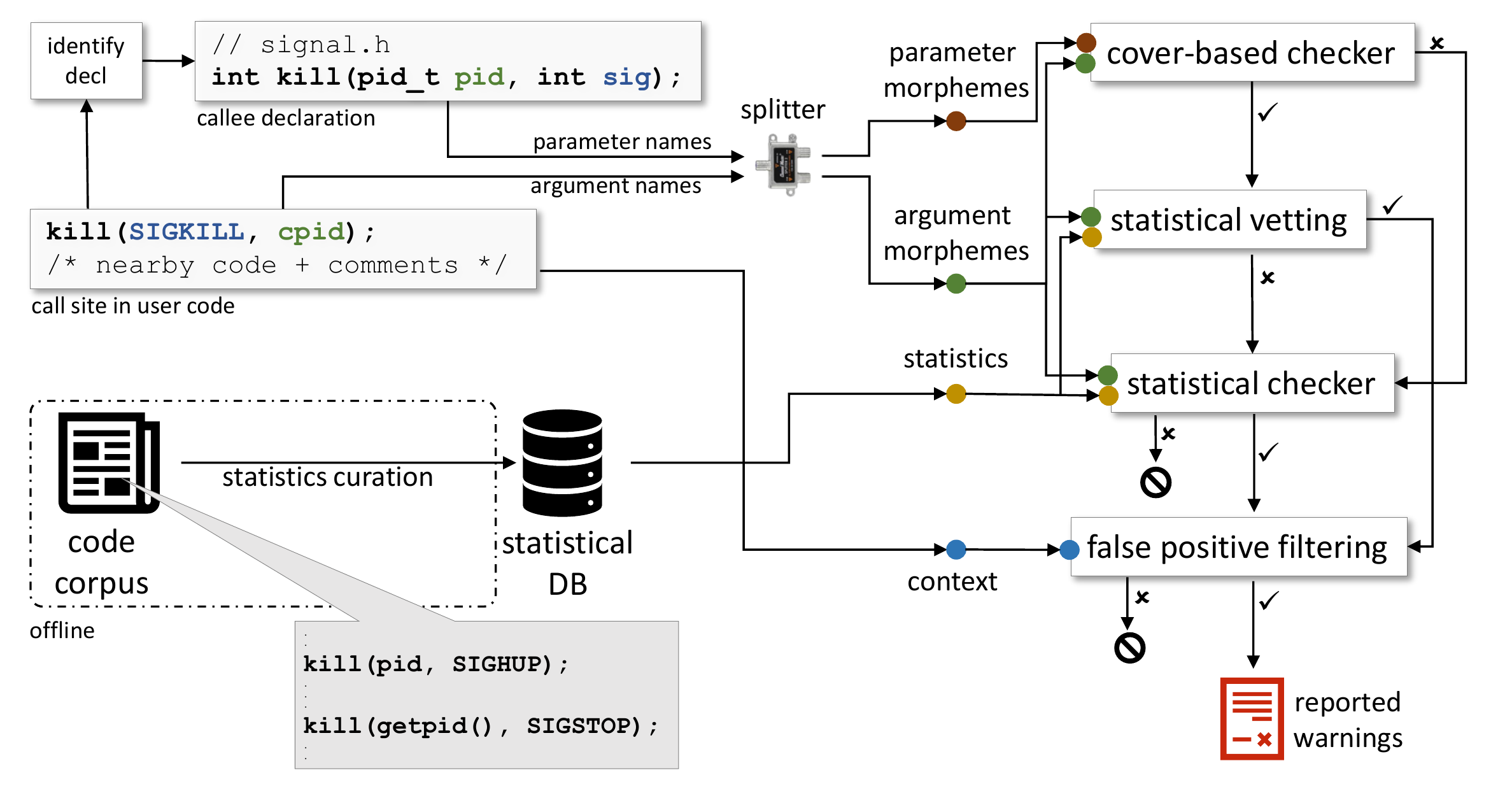}
\caption{\label{fig:overview}High-level overview of \projectname{}. The top left quadrant shows the input to the checker: a call site and its corresponding declaration. The bottom left quadrant shows the offline curation---done once per corpus---of the statistical database, which is available to the checker. The argument names and parameter names corresponding to a call site are split and then processed in four stages, shown in the right half of the diagram.
The sequencing of the four stages is shown using arrows marked with \cmark~(if there may be a warning to report) and \xmark~(if there is no warning to report).
If a swapped-argument error at the call site is still suspected after the false-positive filtering stage, a warning is reported as shown in the bottom right quadrant.}
\end{figure*}

In this section, we give an overview of \projectname{}. We include a number of forward references to \S\ref{label:design} that contain further details on relevant algorithms, techniques, and heuristics.

Figure \ref{fig:overview} is an overview of \projectname{}, featuring the bug in Listing \ref{bug:sigkill}.
The top left quadrant shows the input to \projectname{}: the call site being checked, and the corresponding function declaration. The function declaration is an optional input---if it is not available, then the cover-based checker is skipped. 
Given a call site, we extract names (\S\ref{subsec:names}): from the argument expressions at the call site and from the parameters in the callee function declaration (if available).
Next, we split (\S\ref{subsec:splitting}) both argument and parameter names into morphemes.

Offline, we use a large corpus of code to compute a \emph{statistical database} (\S\ref{subsec:database}), shown in the bottom left quadrant of Figure \ref{fig:overview}. The database is a key-value store, where keys are triples consisting of a function name, argument position, and morpheme. The values are \emph{weight}s indicating the number of projects in the corpus where the morpheme appears in calls to that function, at that argument position. Informally, the weight reflects the number of human programmer communities who considered that the morpheme is appropriate to use at the given argument position for that function.

The right-hand portion of the diagram shows the \projectname{} pipeline of four stages. First, we compare the parameter morphemes and argument morphemes in the \emph{cover-based checker} (\S\ref{subsec:cover}). This stage does not need the statistical database, but it does require the function declaration with parameter names.

If the cover-based checker finds a suspected error, \projectname{} uses the statistical database to perform further vetting (\S\ref{subsec:vetting}) of the warning.
The vetting rules out false positives due to usage patterns for certain functions where seemingly-swapped argument orderings are not rare, indicating that they could have a genuine and intentional use case. 
If we did report such warnings, there could be a lot of false positives due to function-specific patterns adopted by developers. 
If the suspected error passes the vetting step, we move on to the false-positive filtering stage described further below.

\begin{lstlisting}[language=C,escapechar=\#, caption=Example call where seemingly-swapped arguments are \\statistically not rare---thus indicating a likely intentional swap.,label=example:vetting]
// declaration in GStreamer
guint64 gst_util_uint64_scale (guint64 val, guint64 #\dashuline{num}#, guint64 #\uwave{denom}#);

// use in gst-plugins
diff = gst_util_uint64_scale_int (diff, #\uwave{denom\_rate}#, #\dashuline{num\_rate}#);
\end{lstlisting}
\autoref{example:vetting} shows a function declaration from GStreamer~\cite{GStreamer} and a callsite with a likely intentional swap. 
It is statistically not rare to call \texttt{gst\_util\_uint64\_scale\_int} with the morphemes \texttt{denom} and \texttt{num} in the second and third argument positions respectively, i.e., in a swapped order based on parameter names, possibly as a shortcut for computing the reciprocal of the fraction. We discard such warnings.

If the cover-based checker and the statistical vetting do not find any errors at a call site, or are not applicable (as in \autoref{bug:statistical}), \projectname{} runs the statistical checker (\S\ref{subsec:statistical}) to look for other evidence of potential errors using data from the statistical database. Intuitively, we look for pairs of morphemes that appear at two argument positions at the call site, with the property that, statistically, each morpheme is significantly more common at the other's position than at its own. Hypothetically, suppose the cover-based checker was not able to identify the error shown in \autoref{fig:overview}.
The statistical checker gives \projectname{} another chance to catch the error: the statistical database suggests that the \texttt{pid} morpheme is often used in the first position at a \texttt{kill} call site, and the \texttt{sig} morpheme is often used in the second position.
This statistical data suggests that the morphemes at this call site may have been swapped.
Note that both the cover-based checker and the statistical checker could have identified the same error: we quantify how often such an overlap occurs in (\autoref{fig:overlap-tp}, \S \ref{subsec:eval-configs}). 

The final stage for all candidate warnings is false-positive filtering (\S\ref{subsec:fpfiltering}): it applies various heuristics to distinguish between intentional and mistaken swaps.
For example, consider \autoref{bug:iconv}, which presents a false-positive finding in GrafX2~\cite{GrafX2}, filtered out by \projectname{}. 
The second call to \texttt{iconv\_open} at line \texttt{6} uses argument names that appear to be swapped; however, there is a call to the same function with the arguments in the canonical order on the preceding line.
If the programmer calls the function ``both ways'', with calls in close proximity to each other, it is likely that both usages are deliberate (because it undermines the theory that the swap was due to not knowing the correct order). 
We use several false-positive filtering heuristics, including one motivated by the call pattern in~\autoref{bug:iconv}.

\begin{lstlisting}[language=C,escapechar=\#,caption=The candidate warning on line \texttt{6} is ruled out in the false-\\positive filtering stage because of the nearby correct call on line \texttt{5}.,label=bug:iconv]
// declaration in iconv.h
iconv_t iconv_open(const char *#\dashuline{tocode}#, const char *#\uwave{fromcode}#);

// use in grafx2
cd = iconv_open(TOCODE, FROMCODE); // From UTF8 to ANSI
cd_inv = iconv_open(#\uwave{FROMCODE}#, #\dashuline{TOCODE}#); // From ANSI to UTF8
\end{lstlisting}

Throughout our pipeline, we use techniques to minimize noise and maximize signal in the natural language information.
One such technique is comparing morphemes to each other using a similarity metric that takes into account abbreviations (\S\ref{subsec:abbr-synonyms}): for example, \texttt{msg} is a common abbreviation for \texttt{message}.
Another technique is to remove morphemes that are common to pairs of argument names being checked, such as \texttt{remote} in~\autoref{bug:common-morpheme}. 
Removal of the common morpheme allows \projectname{} to detect this bug in BoNeSi~\cite{Bonesi}.

\begin{lstlisting}[language=C,escapechar=\#, caption=Removing the common morpheme \texttt{remote} from the \\argument names at line \texttt{5} clarifies their relationships with the \\parameter names: the third and fourth arguments appear to be swapped.,label=bug:common-morpheme]
// declaration in libnet
libnet_ptag_t libnet_build_tcp(uint16_t sp, uint16_t dp, uint32_t #\dashuline{seq}#, uint32_t #\uwave{ack}#, /* 9 more parameters ... */);

// use in bonesi
if(libnet_build_tcp(origSrcPort, dstPort, #\uwave{remoteAck}#, #\dashuline{remoteSeq}#,  /* 9 more arguments ... */)==-1) { /% ... %/
\end{lstlisting}

In summary, \projectname{} uses a hybrid approach based on both non-statistical and statistical techniques to detect and confirm swapped-argument errors.

\section{Design and Implementation}
\label{label:design}
This section provides details about specific stages, algorithms, and heuristics of \projectname{}.

\subsection{Extracting name information}
\label{subsec:names}
\projectname{} begins by extracting \emph{names} from argument expressions at call sites.
If the corresponding declarations are available and include parameter names, those names are also extracted. We modeled our name extraction on DeepBugs~\cite{Pradel2018}, with adaptations to C and C++.
For an abstract syntax tree node $n$, we extract a string \textit{name}($n$), where possible, as follows.
\begin{itemize}
\item If $n$ is an identifier, return its name.
\item If $n$ is a non-string literal, return a string representation of its value.
\item If $n$ is \texttt{this}, return ``this''.
\item If $n$ is \texttt{(}$m$\texttt{)}, return \textit{name}($m$).
\item If $n$ is one of \texttt{++}$m$, \texttt{--}$m$, $m$\texttt{++}, or $m$\texttt{--}, return \textit{name}($m$).
\item If $n$ is $\otimes m$, where $\otimes \in$ \{\texttt{\&}, \texttt{+}, \texttt{-}, \texttt{*} \}, return \textit{name}($m$).
\item If $n$ is \texttt{sizeof}($m$), return ``sizeof''.
\item If $n$ is a cast or explicit type conversion, return the name of the operand.
\item If $n$ is $l$\texttt{.}$m$, $l$\texttt{->}$m$, or $l$\texttt{::}$m$, return \textit{name}($m$).
\item If $n$ is $l$\texttt{[}$m$\texttt{]}, return \textit{name}($l$).
\item If $n$ is a call $l$\texttt{.}$m$\texttt{(}\ldots\texttt{)} or $m$\texttt{(}\ldots \texttt{)}, return \textit{name}($m$).
\item If $n$ is a macro identifier, return the macro name.
\item In all other cases, return nothing.
\end{itemize}

To handle C/C++ macros, we use information from the preprocessor input rather than the parser input (which is the preprocessor output), which often allows us to operate on more meaningful symbolic names.
If an entire function call is a result of a macro expansion, or if it is a virtual function call, we skip collecting names from that call site.

\comment{
One area where the adaptation was nontrivial is handling preprocessor macros. When computing a name for an expression that comes from a macro expansion, we do the following, assuming accurate information about the macro expansion is available to our CodeSonar~\cite{CodeSonar} analysis. We start from the final source location after all macro expansion has been performed. We ``peel back'' layers of macros until there are none left, or the next expansion location also encompasses the location of the overall call expression. If that macro covers the entire source range of the expression, we return the macro identifier. Some examples of our macro handling are below.

\begin{lstlisting}[language=C,caption=Macro example 1 - argument name is \texttt{errno},label=macroexample1]
#define FOO(arg) my_function(arg)
#define errno (*__error())
FOO(errno)
\end{lstlisting}

\begin{lstlisting}[language=C,caption=Macro example 2 - extracted name is \texttt{WSIGNALED},label=macroexample2]
#define FOO(arg) my_function(arg)
#define errno (*__error())
#define WSIGNALED(x) (x >> 8)
FOO(WSIGNALED(errno))
\end{lstlisting}

\begin{lstlisting}[language=C,caption=Macro name example 3 - argument has no name,label=macroexample3]
#define errno (*__error())
my_function(errno | 1);
\end{lstlisting}

}

\subsection{Splitting names into morphemes}
\label{subsec:splitting}
We split argument names for the input call site to be checked, and parameter names for function declarations. 
We also split argument names when building the statistical database (\ref{subsec:database}).
Our prototype uses the Ronin~\cite{Hucka2018} identifier-splitting algorithm.  
Ronin is an extension of the Samurai~\cite{Enslen2009} algorithm, and uses a global table of token frequencies.
Additionally, during splitting, we drop very common morphemes like ``\texttt{get}'', ``\texttt{set}'', ``\texttt{i}'', ``\texttt{j}'', etc.

\comment{
We first build a global frequency table \textit{GFT} over the entire corpus, by counting the occurrences of argument and parameter names.
This operation is performed offline, and we expect it to be performed infrequently.
We need to redo this operation if we expect a significant change in the name distribution in a corpus.
We set frequency thresholds for the number of occurrences of name: if a name does not appear more than the set frequency threshold, then it is purged from the \textit{GFT}.
Thus, infrequently used names do not affect the splitting process.
The frequency thresholds are set per name length---shorter names have higher thresholds.

The intuition behind using \textit{GFT} for splitting names is that programmers are likely to construct compound names (like \texttt{SIGKILL}) based on simpler names (like \texttt{sig} and \texttt{kill}) that already occur frequently in the corpus.
Additionally, we make use of an English dictionary \textit{ED} in the splitting process.

To split a name into its constituent morphemes, we apply the following steps:
\begin{enumerate}
    \item We split a name on underscore and numeric characters, and when the letters in the name change from lowercase to uppercase. The resulting set of tokens is $S_1$.
    \item Tokens $s \in S_1$ are converted to lowercase, to create $S_2$.
    \item Any token $s \in S_2$ that is present in the dictionary \textit{ED} or is of length less than four, is moved to $S_3'$. Tokens in $S_3'$ are not split further. The rest of the tokens from $S_2$ are moved into $S_3$ for further processing.
    \item We try to split tokens $s \in S_3$ further, in a greedy manner (with backtracking, limited to three attempts). We try to find the longest sub-token present in $s$, that is also present in \textit{GFT} or \textit{ED}, and then recursively split the remainder. We consider a split of $s$ to be  viable if all the tokens from the split are one of: (a) a single character, (b) present in \textit{GFT}, or (c) present in \textit{ED}. Furthermore, for a split to be considered viable, there should not be more than two tokens of length two or shorter, or no more than four tokens of length three or shorter.
    As a result of these machinations, each token $s \in S_3$ could be split in at most three different viable ways. The different viable split outcomes for a token $s$ are scored, and the outcome with the highest score is picked. If $t_1, \ldots, t_n$ is a viable split of $s$, then its score\footnote{As an example, \texttt{threadp} has two viable splits: \{\texttt{th}, \texttt{read}, \texttt{p}\} and \{\texttt{thread}, \texttt{p}\}. The latter viable split has a higher score and is picked.} is given by $\sum\limits_{i=1}^{n} \mathit{GFT}(t_i) * |t_i|^2$. If $t_i$ is missing from the \textit{GFT}, \textit{GFT}($t_i$) is defined to be $0$. $|t_i|$ is the length of $t_i$. All the newly obtained tokens from $S_3$ based on this step are placed in $S_4$.
    \item The constituent morphemes are $S_3' \cup S_4$.
\end{enumerate}
}
\subsection{Morpheme similarity metric}
\label{subsec:abbr-synonyms}
Computing the \emph{similarity} between two morphemes is a fundamental operation in \projectname{}.
We define a similarity metric $\sim$ to quantify the degree of correspondence between two morphemes while allowing for abbreviations.
If two morphemes do not have the same first character, their $\sim$ value is zero. 
Otherwise, their $\sim$ value is computed by applying a penalty for each character that must be deleted from a morpheme in order for the resulting strings to contain the same characters in the same order. 
The penalty is lower for vowels than for consonants, decreases toward the end of the string, and is zero for a final ``\texttt{s}'' (to account for singular/plural forms).
Our penalty for missing characters is normalized by the length of the morphemes, so in longer morphemes we allow for more missing characters while still maintaining a high similarity. 
We say two morphemes are sufficiently similar for a particular purpose if the value of $\sim$ is greater than a context-specific threshold. 
Note that $\sim$ can be naturally extended to be aware of \emph{synonyms}; the end of \S\ref{subsec:morpheme-reasoning} presents a brief discussion of such an experimental extension. 

\comment{
In some cases, we use this similarity metric directly in downstream calculations. In others, we make a binary determination on whether two morphemes $m_1$ and $m_2$ are sufficiently similar, based on a specific threshold.}

\subsection{Computing a statistical database}
\label{subsec:database}
The statistical database is keyed by triples consisting of a function name $f$, an argument position $i$, and a morpheme $m$. For each such triple, it contains a weight $w(f, m, i)$. The weight captures the number of projects in the corpus where morpheme $m$ appears at position $i$ in a call site for $f$. 

For a given function $f$, morpheme $m$, and argument positions $i$ and $j$, we can use the weights in the database to compute a numerical \emph{relative frequency} score:
\begin{align*}
\psi(f, m, i, j) = \frac{w(f, m, i)}{w(f, m, j)}
\end{align*}
This score attempts to quantify how much more common the morpheme $m$ is at argument position $i$ than at argument position $j$ at call sites to $f$. 
In the remainder of the paper, we will sometimes use the notation $\psi(m,i,j)$, omitting the function $f$ when it is clear from context.
When we build the database, we use the splitting techniques described in \ref{subsec:splitting}, and we eliminate common morphemes that appear in all argument positions at a call site. 

\subsection{Cover-based checker}
\label{subsec:cover}

The cover-based checker detects swapped-argument errors if the morphemes in the argument names better ``cover'' the morphemes in the parameter names when argument positions are swapped at a call site.
This checker is skipped if a function declaration lacks parameter names.

After splitting the parameter and argument names into morphemes, we proceed pairwise, for each pair of argument positions $i$ and $j$. In the rest of this paper, $A_i$ and $A_j$ denote the sets of argument morphemes at positions $i$ and $j$ respectively;  $P_i$ and $P_j$ denote the sets of parameter morphemes at corresponding positions.

First, we eliminate any morphemes common to $A_i$ and $A_j$, and similarly for $P_i$ and $P_j$, to handle cases like Listing \ref{bug:common-morpheme}.
If this elimination leaves any of $A_i$, $A_j$, $P_i$, or $P_j$ empty, the cover checker does not proceed any further.

Next, we compute the quality of the match, or  ``cover'', from a set of argument morphemes to a set of parameter morphemes. We run this computation for the original order, computing how well $A_i$ covers $P_i$ and $A_j$ covers $P_j$, and for the ``swapped'' order, i.e., how well $A_i$ covers $P_j$ and $A_j$ covers $P_i$.

Informally, a set of argument morphemes cover a set of parameter morphemes if every parameter morpheme is sufficiently similar to (using the metric $\sim$ described in \ref{subsec:abbr-synonyms}) at least one argument morpheme.
This relation is asymmetric: it is possible to have argument morphemes that are not similar to any parameter morpheme, yet still have coverage. 
However, if a parameter morpheme is not similar to any argument morpheme, then there is no coverage. 

We formalize a notion of coverage $\mathit{C}(A, P)$ for an argument morpheme set $A$ and a parameter morpheme set $P$:
\begin{align*}
\mathit{C}(A, P) = \min_{p \in P}\ \max_{a \in A}\ (a\sim p)
\end{align*}

Our criterion for reporting a swapped-argument warning is based on two thresholds, $\alpha_1$ and $\alpha_2$, empirically determined\footnote{\label{note:empirical}We evaluated the results for a variety of settings and choose the best precision/yield trade-off in our judgment. The space constraints prevent us from providing further details of this process for each threshold. 
} to be $0.5$ and $0.75$ respectively. We produce a candidate warning if and only if both of the following hold:
\begin{align*}
(C(A_i, P_i) < \alpha_1) & \wedge (C(A_j, P_j) < \alpha_1) \\
(C(A_i, P_j) > \alpha_2) & \wedge (C(A_j, P_i) > \alpha_2)
\end{align*}
Informally, we require both sufficiently bad coverage in the current positions and sufficiently good coverage in the swapped positions. 

\begin{figure}
\centering
\includegraphics[scale=0.3]{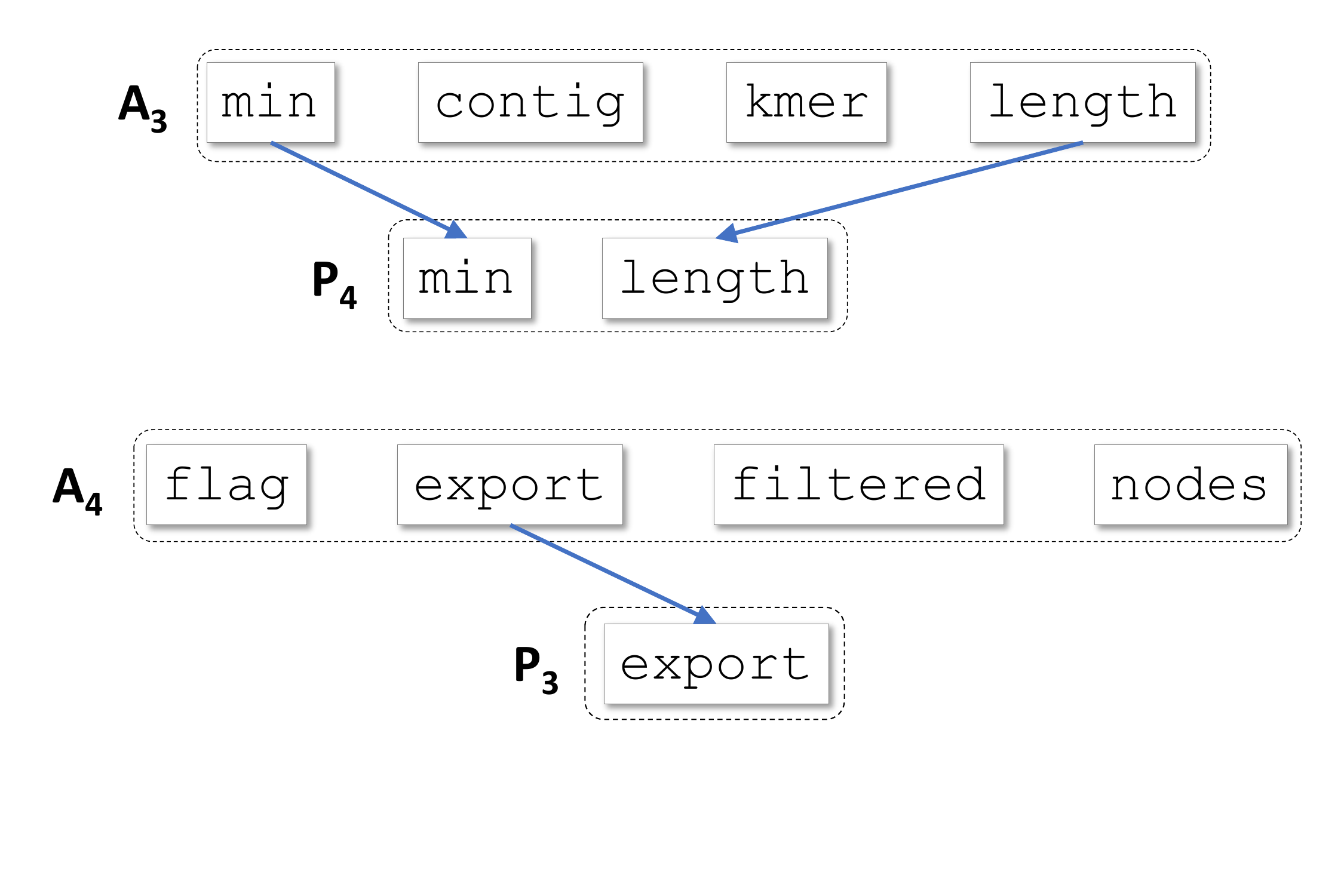}
\vspace{-0.35in}
\caption{\label{fig:cover-match} Depicts the coverage in a swapped position, for the example~\autoref{bug:removeHighCoverageNodes}. 
Here, $\mathit{C}(A_3, P_4) = 1, \mathit{C}(A_4, P_3) = 1, \mathit{C}(A_3, P_3) = 0, \mathit{C}(A_4, P_4) = 0$.  
}
\end{figure}

\begin{lstlisting}[language=C,escapechar=\#,caption=Example bug to showcase the machinery of the cover-\\based checker.,label=bug:removeHighCoverageNodes]
// declaration in a different file
void removeHighCoverageNodes(Graph* graph, double maxCov, boolean #\dashuline{\_export}#, Coordinate #\uwave{minLength}#, /* more params */);

// use in https://github.com/dzerbino/velvet
removeHighCoverageNodes(graph, maxCoverageCutoff, (Coordinate)#\uwave{minContigKmerLength}#, #\dashuline{flagExportFilteredNodes}#, /* more args */ );
\end{lstlisting}
\autoref{bug:removeHighCoverageNodes} shows an example of a bug found using the cover-based checker. \autoref{fig:cover-match} depicts the strong coverage mapping when arguments at position $3$ and $4$ are swapped.

\subsection{Statistical vetting}
\label{subsec:vetting}
We perform statistical vetting when the cover-based checker flags a pair of arguments in positions $i$ and $j$ at a call site as potentially swapped.
We compute $\max_{m \in A_i} \psi(m, i, j)$ and $\max_{m \in A_j} \psi(m, j, i)$.
If either of these quantities exceeds a vetting threshold $\beta$ (empirically determined\footnotemark[4] to be $1$), we conclude that the usage in question is statistically not rare and so do not report a warning. 
\autoref{example:vetting} shows such a case.

\comment{
If these quantities fall under the vetting threshold but above a lower \emph{scoring penalty threshold}, we penalize the warning in our scoring (\ref{subsec:scoring}).
}

\subsection{Statistical checks}
\label{subsec:statistical}

When the cover-based checker does not find a warning at a call site, we run the statistical checker.
It uses the statistical database (\ref{subsec:database}) and the argument names at the call site. 
The statistical checker can run even when the callee declaration does not include parameter names or cannot be retrieved.

If the statistical database does not include statistics for the function called at a call site, the statistical checker is skipped. Otherwise, it considers every possible pair of argument positions $i$ and $j$, and detects instances where two morphemes are likely swapped across those positions using the following approach.
As before, we eliminate common morphemes between $A_i$ and $A_j$.
After such elimination, if either $A_i$ or $A_j$ are empty, we skip the rest of the steps below for $i$ and $j$.

We now look for pairs of argument morphemes $a_i \in A_i$ and $a_j \in A_j$ such that $min(\psi(a_i, j, i), \psi(a_j, i, j)) > \gamma$, for a threshold $\gamma$ (empirically selected\footnotemark[4] to be $5$). Informally, we look at how much more common each morpheme is in the other's position than in its own, and require the lesser of those two ``misplacement'' scores to be greater than $\gamma$.

We also require that $A_i \setminus {a_i} = A_j \setminus {a_j}$, i.e., exactly one morpheme is swapped from the two morpheme sets $A_i$ and $A_j$.
If we find such a pair of morphemes $a_i$ and $a_j$, we perform one more check. We find the morpheme $m$ with the biggest statistical difference in frequency between position $j$ and position $i$, i.e, $m = \argmax_x [w(f,x,j) - w(f,x,i)]$.
We verify that $a_i$ is sufficiently similar to $m$. The intuition is that if morpheme $a_i$ is common in both positions $i$ and $j$, then the likelihood of a swap is lower; we are looking for evidence that moving $a_i$ from $i$ to $j$ would bring the situation \emph{closer} to what is statistically most common. We perform a symmetric check for $a_j$. If both checks pass, we produce a candidate swapped-argument warning involving argument positions $i$ and $j$ and proceed with further false-positive reduction.
Note that the requirements and checks in this paragraph could be relaxed to potentially catch more bugs, however, at the expense of increased false positive rates.

\subsection{False-positive filtering}
\label{subsec:fpfiltering}
Weeding out intentionally- vs. mistakenly-swapped arguments can be difficult.
We have developed a collection of heuristics to identify likely intentional swaps.
Some of them come from the literature; we developed others empirically by manually examining \projectname{} warnings, identifying false positives, and formalizing common features of those false positives.
Without false-positive filtering of this nature, the developer experience of using such checkers can be frustrating.
We list our major heuristics below.

{\bf White-list words.} Some words hint that a swap might be intentional, e.g., ``swap'', ``exchange'', ``rotate'', or ``flip''. We expand on the ``nested in reverse'' heuristic~\cite{Rice2017}, and look for such words in the following locations: the name of the callee function, the name of the caller function, nearby conditional expressions (i.e., the last five branches along the current execution path, see \autoref{bug:rotate-fp} from Mate Panel Libs~\cite{MatePanelLibs}), the six immediately-preceding lines of source code (including any comment contents).
We consider presence of such words to be indicative of false positives, and filter out such warnings. 

\begin{lstlisting}[language=C,escapechar=\#,caption={False-positive warning filtered out, because ``rotate'' is in a \\nearby conditional expression.},label=bug:rotate-fp]
// decl in gdk-pixbuf.h    
GdkPixbuf* gdk_pixbuf_new (/* 3 params */, int #\dashuline{width}#, int #\uwave{height}#); 

// use in mate-panel-libs
if (background->rotate_image /* && ..*/
 // .. several lines of code, and nested if
r = gdk_pixbuf_new (/* 3 args */, #\uwave{height}#, #\dashuline{width}#);
\end{lstlisting}



{\bf Swap distance.} We found that on real-world code, all warnings for argument positions $i$ and $j$ with $|i-j| > 2$ were false positives. 
Therefore, we do not report such warnings.

{\bf Geometric code patterns.} In geometric code, it is common to combine swapping of axes with negation of one of two values to achieve various transformations.
We exclude as intentional any apparent swapped-argument calls that negate exactly one of the two arguments involved.

{\bf Type checking.} We eliminate some false positives through simple type checking on the types of the two arguments involved (similar to~\cite{Liu2016,Pradel2013}).
The intuition is that a swap of arguments with incompatible types would have been detected in development via a compiler error.
Even among compatible types, if the swapped order requires more type coercion, that may argue for the correctness of the existing code.

{\bf Nearby declaration.} If the declaration of the callee function is in the same source file as the call site, we consider the warning a false positive. This heuristic is based on our empirical findings, and on the intuition that erroneous swaps happen when the programmer forgets the correct argument order. If the declaration is nearby, the programmer likely is aware of the correct argument order.

{\bf Nearby correct call.} If there are other calls to the same function, but with unswapped arguments, within the same caller function (see Listing~\ref{bug:iconv}), we consider the warning a false positive.
This heuristic (similar to the ``duplicate method calls'' heuristic~\cite{Rice2017}) is based on empirical findings, and on an intuition about reminder proximity similar to that of the previous heuristic.

{\bf Swap is not rare.} If the suspected swap is not an isolated event, but occurs in three or more separate callsites within the same calling function, we consider it a false positive.
Our observation is that anomalies tend to be intentional unless they occur very rarely.
This heuristic causes us not to report cases where a function is called at several callsites consistently with the wrong argument order; however, true positives of this kind are far outweighed by false positives.

\comment{
\subsection{Scoring}
\label{subsec:scoring}
\numberedtodo{RS}{TP 135978.}
}

\section{Evaluation}
\label{sec:eval}

In this section, we present the results of an empirical evaluation of \projectname{} on a large C and C++ code corpus. 
The major research questions we consider are:
\begin{questions}
\item How well does \projectname{} find warnings in real-world open-source code?\label{rq:1}
\item What is the value-add of the different stages in \projectname{}?\label{rq:2}
\item (a) How often are argument and parameter names constructed from multiple morphemes? (b) How many of the true positives found involve multiple morphemes? These questions are aimed at seeking justification for morpheme-level reasoning, instead of operating directly on whole names.\label{rq:3}
\item (a) What effect does the corpus size used for the statistical database have on \projectname{}'s findings? (b) What is the effect of leaving out those projects from the statistical database, on which \projectname{} produces any warnings?\label{rq:4}
\end{questions}
We also provide a discussion of the warnings triaged (\S \ref{subsec:triaged-warnings}) and threats to validity (\S \ref{subsec:threats}) of our work. 

\subsection{Prototype implementation and corpus}
\label{subsec:prototype-corpus} 

We use the commercial static analysis tool CodeSonar~\cite{CodeSonar} to extract name information from call sites and their corresponding declarations, when available. 
We implemented the statistics database computation (\S \ref{subsec:database}) and the \projectname{} checker prototype in Python. 
Our prototype generates warnings in the SARIF format~\cite{SARIF}. 
Such warnings can be imported into an IDE or other SARIF viewers (such as CodeSonar) for manual inspection and triage.
CodeSonar remembers the triage result (i.e., true/false positive), as well as other user annotations, by fingerprinting the warning location.
We found this ability to be helpful for manual construction of ground truth for the evaluation, and during review of the results.

We computed the statistical database using the open-source Fedora 29 source-package repository~\cite{FedoraSRPM}, filtered to include only projects containing C or C++ code. 
We performed additional filtering to reduce duplication and eliminated extremely large projects. 
We successfully processed \corpusProjs{} projects, consisting of about \corpusSize{} lines of code. 
We refer to this set of projects as the SRPM corpus in the rest of this paper. 
The resulting statistical database contains morpheme information for over \statsDBSize{} functions. 

\subsection{Evaluation methodology}

We considered evaluating \projectname{} on both real-world code and on a synthetically-generated dataset with randomly injected swapped arguments. 
We decided against the latter, because it is unclear how to generate a synthetic dataset with a realistic distribution of both erroneous and intentional swaps. 
In practice, intentional swaps far outnumber actual swap errors, so we do not believe that a na\"ive injection approach that disregards them would lead to a realistic dataset; thus, evaluation results on synthetic datasets may not carry over to real-world code. 
Therefore, we decided to conduct an evaluation exclusively on real-world open-source code.
We perform our evaluation on the SRPM corpus, i.e., \corpusSize{} lines of C and C++ code. 
As far as we are aware, our evaluation is the largest (in terms of number of lines of code) for a swapped-argument checker on any programming language~\cite{Rice2017,Pradel2018}; and the largest by far~\cite{Pradel2013} on C/C++. 

A limitation of using real-world code for evaluation is the lack of pre-existing ground truth. 
To obtain a list of true- and false-positive warnings, we ran \projectname{} under different configurations (described in \ref{subsec:eval-configs}) to obtain a total of \num{4141} unique warnings reported on the SRPM corpus.
Of these, we sampled and manually triaged \warningsTriaged{} unique warnings: we marked \triagedTPs{} of these as true positives, and \triagedFPs{} as false positives. 
When \projectname{} is run again on the SRPM corpus under any configuration, a warning reported at a triaged location is recognized and automatically classified as a true or a false positive.
The manual triage task was shared by six experienced developers, some of whom were involved in the development of \projectname{}. 
We applied a conservative triage strategy---marking warnings as true positives if they reflect issues worth raising in a code review (i.e., real bugs or problems worth fixing even if there is no runtime error). 
Otherwise, we marked warnings as false positives.
\autoref{bug:initv6} from OpenVAS libraries~\cite{OpenVAS} shows an example warning marked as a false positive: we suspect that the swap on line \texttt{6} is intentional, because on line \texttt{5}, the format string ``\texttt{src host \%s}'' uses an argument computed from \texttt{dst}. 

\begin{lstlisting}[language=C,escapechar=\#,caption=Warning triaged as a false positive.,label=bug:initv6]
// declaration 
int init_v6_capture_device (struct in6_addr #\dashuline{src}#, struct in6_addr #\uwave{dst}#, char *filter);

// use in openvas-libraries
snprintf (filter, sizeof(filter), "ip6 and src host %s", inet_ntop(AF_INET6, dst, addr, sizeof (addr)));
bpf = init_v6_capture_device (*#\uwave{dst}#, #\dashuline{src}#, filter);
\end{lstlisting}

                	

Because we manually triaged a \emph{sample} of warnings, we use precision and yield as our evaluation metrics. 
Precision is the ratio of the number of true-positive warnings to the total number of warnings. 
Yield is the total number of reported true-positive warnings from our ground-truth dataset. 
Yield is a proxy for recall: since there is no practical way to determine the full set of all swapped-argument errors in the corpus, we cannot determine what percentage of them we have found.

Making \projectname{} practically useful requires balancing precision and yield. 
High precision with low yield leads to few reported warnings; while these warnings are likely to be real problems, many other real problems may be missed. 
Low precision with high yield is also not ideal, because it leads to large numbers of false positives. 
The developer effort to sift through those can cause frustration and reduce adoption.
In a practical tool, scoring and sorting can be used to balance these conflicting concerns. 
By assigning scores to warnings based on their likelihood of being true positives, we can show them to the user in a descending order. 
The user can then decide when the effort of further manual triaging is no longer justified by the likely benefit of discovering an additional true positive.
Scoring the warnings from \projectname{} is an interesting problem that is outside the scope of this paper.

\subsection{Evaluating various stages}
\label{subsec:eval-configs}

As outlined in Section \ref{sec:overview}, \projectname{} involves a multi-stage pipeline, with four stages: (1) cover-based checker, (2) statistical vetting, (3) statistical checker, and (4) false-positive filtering. 
Each of these stages impact precision and/or yield. 
To expose the impact of each stage, we evaluated the precision and yield of \projectname{} in a variety of different configurations. 
In each configuration name, the numbers indicate which of the four stages are enabled.
For example, in \configuration{3}, only stage $3$ is enabled, and the rest of the stages are disabled. 

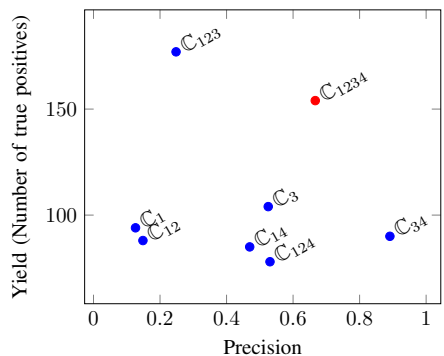
\begin{figure}
  \centering
\begin{tikzpicture}[scale=0.8]
  \begin{axis}[
    enlargelimits=0.2,
    xlabel=Precision,
    ylabel=Yield (Number of true positives),
  ]
      \addplot[
          scatter/classes={a={blue}, b={red}},
          scatter, mark=*, only marks,
          scatter src=explicit symbolic,
          nodes near coords*={\label},
          every node near coord/.append style={rotate=30, anchor=west},
          visualization depends on={value \thisrow{label} \as \label}
      ] table [meta=class] {
          x y class label
          0.1263 94 a \configuration{1}
          0.1486 88 a \configuration{12}
          0.2482 177 a \configuration{123}
          0.6667 154 b \configuration{1234}
          0.5306 78 a \configuration{124}
          0.4696 85 a \configuration{14}
          0.5253 104 a \configuration{3}
          0.8911 90 a \configuration{34}
      };
  \end{axis}
  \end{tikzpicture}
\caption{\label{fig:py-config} Precision vs. yield for various \projectname{} configurations. \configuration{1234} has the best trade-off between precision and yield.}
\end{figure}

Figure~\ref{fig:py-config} shows the precision and yield of \projectname{} for various configurations. 
A few observations are clear: \configuration{1234} shows the best trade-off between precision and yield. 
\configuration{123} has higher yield, but very low precision---it reports an order-of-magnitude more warnings in total than \configuration{1234}.
However, 75\% of the warnings from \configuration{123} are false positives, which justifies our adoption of the false-positive filtering stage to increase precision. 
In fact, all the configurations without stage $4$ enabled (i.e., \configuration{1}, \configuration{12}, \configuration{3}, \configuration{123}) have low precision.

Configurations other than \configuration{1234} and \configuration{123} have much lower yield, suggesting that relying solely on either the cover-based checker or the statistical checker misses several true-positive warnings, justifying our hybrid approach. 
Comparing \configuration{12} vs. \configuration{1}, and \configuration{14} vs. \configuration{124}, we see a small increase in precision traded for a small decrease in yield: thus, cross-checking with the statistical database to perform statistical vetting of the cover-based checker warnings can be useful if a user prefers precision over yield. 

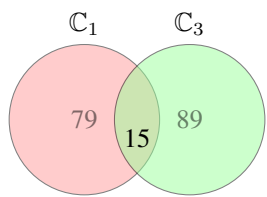
\begin{figure}
  \centering
\begin{tikzpicture}
  \tikzset{venn circle/.style={draw,circle,minimum width=2cm,fill=#1,opacity=0.5}}

  \node [venn circle = red!40!white, label=\configuration{1}] (A) at (0,0) {\ccOnlyTP{}};
  \node [venn circle = green!40!white, label=\configuration{3}] (B) at (0:1.4cm) {\scOnlyTP{}};
  \node[below] at (barycentric cs:A=1/2,B=1/2 ) {\overlapTP{}};   
\end{tikzpicture}  
\caption{\label{fig:overlap-tp} Presents the number of unique true-positive warnings reported by cover-based checker (\configuration{1}) only, statistical checker (\configuration{3}) only, and by both.}
\end{figure}

In summary, these results answer~\ref{rq:1}: with all the four stages enabled, \projectname{} finds \baselineTP{} true-positive warnings\footnote{\configuration{1234} reports \num{402} warnings in total on the SRPM corpus. Of these, \num{231} warnings are present in our manually-triaged ground-truth dataset. Among these \num{231} warnings, \num{154} are true positives.} on the SRPM corpus, with a precision of \baselinePrec{}.
These results also answer~\ref{rq:2}: each of the four stages contribute to either increasing precision or increasing yield, justifying the use of each stage.
\autoref{fig:overlap-tp} shows the overlap in true-positive warnings between the cover-based checker and the statistical checker---both these approaches largely find different sets of true warnings, further bolstering the case to use both of them. 

\subsection{The case for morpheme-level reasoning}
\label{subsec:morpheme-reasoning}
As described in \S\ref{sec:intro}, operating on morphemes instead of whole names is a key  feature of our approach. 
Finding bugs such as those in \autoref{bug:common-morpheme} and \autoref{bug:removeHighCoverageNodes} highly benefit from morpheme-level reasoning: using a string-distance metric on whole names is not a good fit for finding such errors. 
However, if almost all the argument and parameter names provided by developers consist of single morphemes, then the morpheme-level reasoning boils down to whole-name-level reasoning, making morpheme-level reasoning overkill.

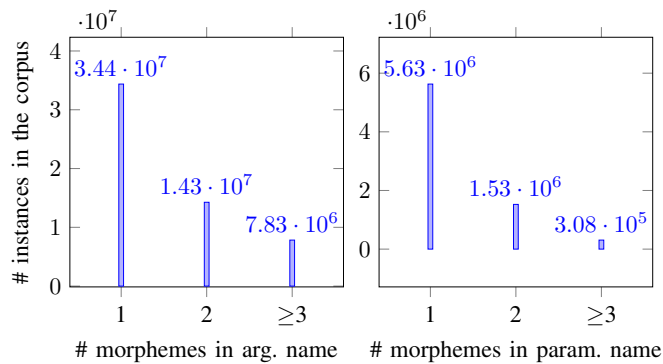
\begin{figure}
  \centering
  \begin{minipage}{0.48\columnwidth}
      \centering
      \begin{tikzpicture}[scale=0.9]
        \begin{axis}[
          xlabel=\# morphemes in arg. name,
          ylabel=\# instances in the corpus,
          xtick=data,
          symbolic x coords={0, 1, 2, $\geq$3},
          enlargelimits=0.3,
          ybar=0.2pt,
          bar width=2pt,
          nodes near coords,
          width=160pt,
          height=150pt
        ]
        \addplot
          coordinates {
            (1,34370562)
            (2,14261637)
            ($\geq$3,7827142)
          };
        \end{axis}
      \end{tikzpicture}
  \end{minipage}\hfill
  \begin{minipage}{0.48\columnwidth}
      \centering
      \begin{tikzpicture}[scale=0.9]
        \begin{axis}[
          xlabel=\# morphemes in param. name,
          xtick=data,
          symbolic x coords={0, 1, 2, $\geq$3},
          enlargelimits=0.3,
          ybar=0.2pt,
          bar width=2pt,
          nodes near coords,
          width=160pt,
          height=150pt
        ]
        \addplot
          coordinates {
            (1,5628433)
            (2,1525033)
            ($\geq$3,307769)
          };
        \end{axis}
      \end{tikzpicture}
  \end{minipage}
  \caption{\label{fig:morphemes-corpus} Frequency distribution of morpheme-set sizes for argument names (left) and parameter names (right) across the SRPM corpus. The y-axis in both the charts (note different scales) indicates frequency; the x-axis provides morpheme-set size.}
\end{figure}

\autoref{fig:morphemes-corpus} answers~\ref{rq:3}(a); it shows how often argument names and parameter names in the SRPM corpus are constructed with different morpheme-set sizes. 
If we cannot extract a name from a call site or a declaration, it does not get counted in this Figure. 
While a majority of both the argument and parameter names chosen by developers are made up of single morphemes, nearly 40\% argument names are constructed from more than one morpheme, lending credibility to our use of morpheme-level reasoning. 



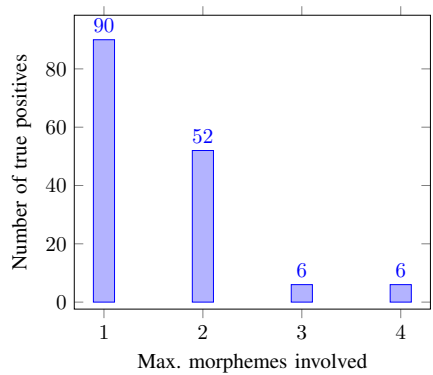
\begin{figure}
  \centering
\begin{tikzpicture}[scale=0.8]
  \begin{axis}[
    xlabel=Max. morphemes involved,
    ylabel=Number of true positives,
    enlargelimits,
    legend pos=north east,
    ybar,
    nodes near coords,
  ]
  \addplot
    coordinates {
      (1,90)
      (2,52)
      (3,6)
      (4,6)
    };
  \end{axis}
  \end{tikzpicture}
  \caption{\label{fig:morpheme-size} Maximum number of morphemes (argument or parameter morphemes in either positions) involved in a reported true positive by \configuration{1234}.}
\end{figure}

Furthermore, in Figure~\ref{fig:morpheme-size}, we plot the frequency distribution of the maximum morphemes involved for all the true-positive warnings reported by \baseline{}: it serves to answer~\ref{rq:3}(b).
For a reported warning, let $i$ and $j$ be the two argument positions involved in the swap, and let $A_i$, $A_j$, $P_i$, and $P_j$ be the argument morpheme and parameter morpheme sets in those two positions respectively.
Then, the maximum morphemes involved in the warning is given by $\max \{ |A_i|, |A_j|, |P_i|, |P_j| \}$. 
Nearly 42\% of true-positive warnings involve names with more than one morpheme, confirming that our morpheme-level reasoning in the different stages are likely useful in identifying real bugs.

\begin{lstlisting}[language=C,escapechar=\#,caption={Example bug identified when \projectname{} considers\\ \texttt{size} and \texttt{count} to be synonyms.},label=bug:synonym]
size_t scm_port_buffer_put (SCM buf, const scm_t_uint8 *src, size_t #\dashuline{count}#, size_t end, size_t #\uwave{avail}#); // decl

// use in guile
scm_port_buffer_put (new_buf, scm_port_buffer_take_pointer (pt->read_buf, cur), #\uwave{avail}#, 0, #\dashuline{c\_size}#);
\end{lstlisting}

Inclusion of \emph{synonym} relationships in the similarity metric $\sim$ would allow morpheme-level reasoning to find even more errors.
For example, if we consider \texttt{size} and \texttt{count} as synonyms, i.e., $\sim(\mathtt{size},\mathtt{count}) = 1$, then \projectname{} finds the error shown in~\autoref{bug:synonym} from Guile~\cite{Guile}. As future work, we want to automatically extract synonyms from code corpora, and extend our morpheme-similarity metric with knowledge of these synonyms.



\subsection{Effect of corpus used for computing the statistical database}

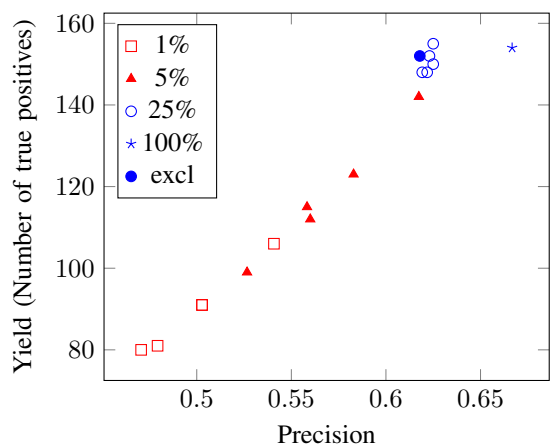
\begin{figure}
  \centering
  \begin{tikzpicture}
    \begin{axis}[
      xlabel=Precision,
      ylabel=Yield (Number of true positives),
      legend pos=north west
      ]
        \addplot[
        scatter/classes={
          g={mark=square,draw=red,fill=red},
          a={mark=triangle*,red},
          c={mark=o,draw=blue,fill=blue},
          f={mark=star,draw=blue},
          e={mark=*,draw=blue,fill=blue}
        },
        scatter,only marks,
        scatter src=explicit symbolic]
      table[x=x,y=y,meta=label]{
      x y label
      0.5408 106 g
      0.623 152 c
      0.6174 142 a
      0.4706 80 g
      0.6218 148 c
      0.56 112 a
      0.5028 91 g
      0.6192 148 c
      0.5583 115 a
      0.4793 81 g
      0.625 155 c
      0.5266 99 a
      0.5028 91 g
      0.625 150 c
      0.5829 123 a
      0.6179 152 e
      0.6667 154 f
      };
      \legend{1\%,5\%,25\%,100\%,excl}
    \end{axis}
  \end{tikzpicture}
\caption{\label{fig:corpus-size-effect}Precision and yield reported when \baseline{} is run on the SRPM corpus with the statistical database computed from a random subset (i.e., 1\%, 5\%, 25\%, and 100\%) of the projects in the SRPM corpus. Each random selection is made five times: due to overlapping points, fewer than five points per subset size may be visible. The legend ``excl'' refers to using a statistical database that was computed without those \num{239} projects that had at least one warning reported by \baseline{}. Note that the origin in this chart is not at zero precision and zero yield.}
\end{figure}

In \autoref{fig:corpus-size-effect}, we take different randomly-chosen subsets (1\%, 5\%, 25\%) of the SRPM corpus to compute the statistical database, and present the results of running \baseline{} on the SRPM corpus with these different statistical databases. 
Each random subset is computed five times. 
This plot serves to answer~\ref{rq:4}(a), showing the effect of the corpus size used for the statistical database on \projectname{}'s findings. All the five random trials with 1\% of the corpus, and most of the five random trials with 5\% of the corpus, are in the bottom-left quadrant (low precision and low yield). 
However, all the rest of random subsets are in the top-right quadrant (high precision and high yield), suggesting that statistical database computation from relatively small subsets of the SRPM corpus (i.e., sometimes even just 5\%) can provide \emph{most} of the precision and yield gains, compared to using the entire corpus. 
Note that the cover-based checker has only a weak, second-order dependency on the statistical database.
If no statistics are available to vet its results, the yield can only  increase, with a decrease in precision.

Furthermore, when computing the statistical database, if we exclude all the \num{239} projects in the SRPM corpus that \configuration{1234} reports a warning on, the precision and recall are not affected much (see legend ``excl'', both precision and recall are only slightly lower than when using the entire corpus). 
This observation answers~\ref{rq:4}(b) and confirms that the analysis is not over-fit to the specific projects within which it is reporting warnings.

\subsection{Discussion of triaged warnings}
\label{subsec:triaged-warnings}



\begin{figure}
  \centering
\begin{tikzpicture}[scale=0.8]
  \begin{axis}[
    xlabel=Number of arguments at a callsite,
    ylabel=Probability of a true positive warning,
    xtick=data,
    symbolic x coords={2, 3, 4, 5, 6, 7, $\geq$8},
    enlargelimits=0.2,
    legend pos=north east,
    ybar=0.2pt,
    bar width=2pt,
    width=300pt,
    height=180pt,
    nodes near coords,
  ]
  \addplot
    coordinates {
      (2, 1.3646702047005307e-05)
      (3, 2.850754611514813e-05)
      (4, 1.3173652694610779e-05)
      (5, 2.6699029126213593e-05)
      (6, 2.4038461538461542e-05)
      (7, 4.901960784313726e-05)
      ($\geq$8, 4.387098032142805e-05)
    };
  \end{axis}
  \end{tikzpicture}
  \caption{\label{fig:args-count-tp}Probability of a true positive warning occurring at a call site with $n$ arguments (for $n = 2, \ldots, \geq 8$). 
  We computed these probability values in a manner similar to Figure 11 in \cite{Rice2017}.}
\end{figure}
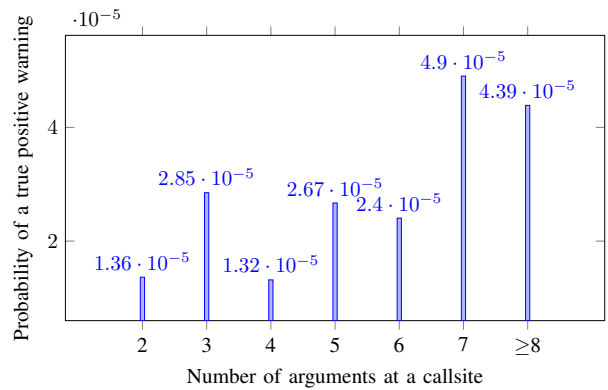

\autoref{fig:args-count-tp} compares the relative probabilities of a true-positive warning occurring at a call site with a given number of arguments. 
To compute these probabilities, we leave out call sites with less than two arguments, because it would not affect the relative probabilities shown here. 
We computed these probabilities as described for Figure 11 in~\cite{Rice2017}. 
In contrast to the probability distribution for Java programs~\cite{Rice2017}, we found that probability of true-positive warnings given a call site with two or three arguments is comparable to the probabilities of those with higher number of arguments. 
One possible explanation for the differences in the probability distributions could be that because C and C++ programs are ``weakly typed'', it allows more room for confusion in ordering arguments, even when involving call sites with few arguments. 


In our triage of \projectname{} warnings, we found a variety of reasons for  false-positive warnings. 
Some example reasons include: incorrect splitting of names into morphemes, incorrectly detected abbreviations, function-specific patterns that statistical vetting is not able to pick up on, poor naming decisions by the developer, patterns that are rare but not incorrect, and names that don't carry much meaning (see ~\autoref{bug:xscreensaver} from XScreenSaver~\cite{XScreenSaver}).

\begin{lstlisting}[language=C,escapechar=\#,caption={Likely false-positive warning: co-ordinate system could be\\ altered in OpenGL.},label=bug:xscreensaver]
// decl in OpenGL
void glVertex3f(GLfloat x, GLfloat #\dashuline{y}#, GLfloat #\uwave{z}#);

// use in xscreensaver
glVertex3f (x, #\uwave{z}#,#\dashuline{ y}#);
\end{lstlisting}







\subsection{Threats to validity}
\label{subsec:threats}

Our techniques assume English names; it is unclear how much of our work is applicable to non-English names. 

Our statistical database is derived from a mature open-source code corpus for the Linux platform, and this particular corpus may have good coding patterns, which is likely beneficial. 
However, we may have higher yield on projects that are less mature or yet-to-be released. 
Similar to a lot of work in this research area---where patterns are mined from code---our statistical vetting and the statistical checker makes the assumption that ``most code is correct''.
However, in specific domains, this assumption may not hold~\cite{Egele2013}. 
We give more importance to statistical patterns that occur across \emph{several} projects, which may help assuage some concerns about our assumption.
A possible area for improvement would be to recognize similar code in different projects and discount the statistics for occurrences across multiple similar projects. 
We do this deduplication (\S \ref{subsec:prototype-corpus}) at the granularity of entire files, which ignores many other forms of code duplication.

We expect our work to be applicable for several popular programming languages that support position-based arguments, other than C and C++; however, techniques in \projectname{} may not be useful for programs written in languages with keyword arguments, such as Smalltalk and Objective-C. 

Finally, many of the warnings were triaged by people who developed \projectname{}, which could have caused some bias in labeling warnings. 
One possible source of such bias is in the sampling of the warnings to triage. 
If the validity of a warning is difficult to ascertain, the triager may skip it and look for an easier one. 
However, difficult-to-triage warnings are more likely to be false positives, so skipping these would bias the triaged results toward more true positives.

\section{Related Work}
\label{sec:related}

In this section we discuss closely related previous work.

\subsection{Matching argument and parameter names}

The idea of detecting swapped-argument errors using mismatches between argument names and parameter names has been studied before~\cite{Pradel2013,Liu2016,Rice2017}. 
Of these works, Rice et al.\cite{Rice2017} have the most extensive real-world evaluation (run on 200 million lines of proprietary code, and 10 million lines of open-source code).

They detect incorrectly-ordered arguments at call sites in Java programs, and their work is most similar to our cover-based checker. 
They use string-similarity metrics on whole names to detect mismatched correspondences between arguments and parameters, whereas our cover-based checker performs morpheme-level reasoning. 
We believe that the cover-based checker is a better approach because it picks relevant signals from names being compared.
Comparing whole names using a string distance is akin to comparing two whole sentences using a string distance: there is fundamental impedance mismatch; whereas using a cover-based checker is akin to comparing two sentences based on the words contained in them.
Our approach is also readily extended to other morpheme-similarity measures, including considering synonymous morphemes to be equivalent.
Their work will miss reporting bugs if parameter names are not available or not useful, whereas our hybrid approach can still report bugs in such cases based on mined statistical patterns.
Their work will report false positives if there are function-specific anti-patterns that developers use (such as \autoref{example:vetting}), whereas we can filter out such warnings using statistical vetting. 


\subsection{Learning from code}
With the increased availability of large amounts of code, learning models of ``correct'' code from existing programs and detecting \emph{anomalies as bugs}~\cite{Engler2001} has been gaining popularity~\cite{Ramanathan2007,Bian2019,Murali2017,Perl2015,Li2018}; none of these~\cite{Engler2001,Ramanathan2007,Bian2019,Murali2017,Perl2015,Li2018} are mining name information for detecting swapped-arguments errors. 
We discuss our work in contrast to two such closely related works: DeepBugs~\cite{Pradel2018} and APISan~\cite{Yun2016}. 

\textbf{DeepBugs} detects swapped-argument errors using a machine learning approach: they seed a corpus of programs with artificial \emph{likely} swapped-arguments errors, and train a classifier to distinguish the artificial code from the unmodified real code. 
Because the real code is expected to have very few swapped-arguments errors, their hypothesis is that the classifier learns to identify swapped-arguments errors in real code.
They apply their technique to JavaScript programs, with a corpus of $68$ million lines of code.
Their work is most similar to our statistical checker. 

Their artificial seeding of swapped arguments in a corpus does not distinguish between intentional and unintentional swaps, and therefore, their classifier is unlikely to learn such a distinction.
Determining whether a swap is intentional or not requires considering the surrounding code and context (e.g., preceding source text, conditionals, caller function), but such information is not taken into account by DeepBugs. 

DeepBugs only considers swaps between the first two arguments at a call site, whereas we consider swaps between all pairs of arguments.
DeepBugs requires a lot of training data; and it only reports warnings when the whole function name and the whole argument names (at first two positions) at a call site are all present in the top \num{10000} vocabulary of names.
DeepBugs reasons at the whole-name level, so call sites with less-frequently occurring whole argument names (that could be made of frequently-occurring morphemes) and function names are not even considered. 
Our morpheme-level reasoning boosts the signal present in name data for the statistical checker, and our hybrid approach can find bugs even when there is no statistical data available for a particular function.

Being able to explain why a warning is reported is an essential element for adoption. 
Explaining why DeepBugs predicted a call site to be buggy is hard~\cite{Rudin2019,XAI}.
In contrast, our approach provides straightforward algorithmic explanations for each finding.

\textbf{APISan} detects various classes of errors by computing a statistical database of function-usage characteristics, and then finding anomalous patterns in the database. 
The characteristics they extract from arguments at a call site do not pertain to argument ``names''.
Instead, they focus on extracting and statistically reasoning about traditional semantic relations between argument values.

\comment{
\section{Future Work}
\label{sec:future}
\fixup{TODO: brief and included if there is space.}
\begin{outline}
    \1 Evaluation on different programming languages. 
    \1 Suggesting fixes for the warnings.
    \1 Experimenting with more sophisticated splitting and abbreviation detection techniques. 
\end{outline}    
}

\section{Conclusion}
\label{sec:conclusion}


In this paper, we have presented \projectname{}, a technique to find mistakenly-swapped arguments at call sites. 
\projectname{} exploits ``big code'' and carefully combines four stages (cover-based checker, statistical vetting, statistical checker, and false-positive filtering) to balance the precision and yield of the findings.

\section*{Acknowledgments}
This material is based on research sponsored by the Department of Homeland Security (DHS) Office of Procurement Operations, S\&T acquisition Division via contract number 70RSAT19C00000056. The views and conclusions contained herein are those of the authors and should not be interpreted as necessarily representing the official policies or endorsements, either expressed or implied, of the DHS.
We would like to thank Amy Gale for her feedback on our work.

\clearpage

\bibliographystyle{IEEEtran}
\bibliography{paper}

\end{document}